\documentclass[12pt]{iopart}

\usepackage{graphicx}% Include figure files
\usepackage{dcolumn}% Align table columns on decimal point
\usepackage{bm}% bold math
\usepackage{grfext}% fixes epstopdf options
\usepackage[colorlinks=true,citecolor=blue,linkcolor=blue,urlcolor=blue]{hyperref}
\usepackage{xcolor}

\newcommand{\nof}{n}
\newcommand{\binomial}[2]{{{#1} \choose{ #2}}}

\begin{document}

\title{The Fragmentation Kernel in Multinary/Multicomponent Fragmentation}
\author{Themis Matsoukas}

\address{Department of Chemical Engineering, Pennsylvania State University}
\ead{txm11@psu.edu}
\vspace{10pt}
\begin{indented}
\item[]\today
\end{indented}

\begin{abstract}
The fragmentation equation is commonly expressed in terms of two functions, the rate of fragmentation and the mean number of fragments.  In the case of binary fragmentation an alternative  description is possible based on the fragmentation kernel, a function from which the rate of fragmentation and the mean distribution of fragments can be obtained. We extend the fragmentation kernel to multinary/multicomponent fragmentation and derive expressions for certain special cases of random and non random fragmentation. 

\end{abstract}

%\maketitle

%--------------------------------------------------------------------------*
\section{Introduction}
Ziff and coworkers \cite{Ziff:JPMG85,Ziff:Macromolecules:1986,Ziff:JPMG91,Ziff:JPMG92} popularized the treatment of binary fragmentation in terms of the fragmentation kernel, a single function of two variables, the masses in the fragment pair produced, that completely describes the process.  The traditional  approach uses two separate functions, the fragmentation rate as a function  of cluster mass  and the distribution of fragments it produces. Both are obtained from the fragmentation kernel, a property that endows the kernel with the status of a fundamental function from which all other properties of fragmentation can be derived.  The use of a single function makes Ziff's treatment of fragmentation more concise, but also more symmetric to aggregation. Binary aggregation and binary fragmentation have a reverse relationship to each other and the symmetry between the two becomes clear only when both processes are formulated in terms of a kernel function \cite{Matsoukas:Entropy:2022}. 
 
As the generating function for all properties in fragmentation, the kernel ought to serve as the starting point for constructing general fragmentation models. It is not obvious, however, how the kernel should be extended when multiple fragments are involved, much less so if multiple components are also present. The purpose of this paper is to formulate the fragmentation kernel as a single generating function of all relevant properties in the general case of multinary multicomponent fragmentation.  In doing so we will be drawing upon the probabilistic  nature of fragmentation.

%--------------------------------------------------------------------------*
\subsection{The Breakup kernel in binary fragmentation}
To gain insight we begin with a review of the binary case. The fragmentation kernel $F_{i,j}$ in binary fragmentation characterizes the break up event
\begin{equation}
  (i+j) \stackrel{F_{i,j}}{\longrightarrow} (i) + (j),
\end{equation}
and gives the rate at which parent mass $i+j$ produces the ordered pair $(i,j)$. It satisfies the symmetry relationship
\begin{equation}
  F_{i,j} = F_{j,i} ,
\end{equation}
which states that all permutations of a pair are produced with the same rate. The fragmentation rate of mass $v$ is the total rate at which mass $v$ produces fragments:
\begin{equation}
\label{ak-bin}
  a(v) = \sum_{i=1}^{v-1} F_{v-i,i} .
\end{equation}
The number of fragments $b_{i|v}$ of size $i$  produced from parent mass $v$ is the rate of production of fragments of that size normalized by the fragmentation  rate of of mass $v$:
\begin{equation}
\label{bik-bin}
  b_{i|v} 
  = \frac{F_{v-i,i}+F_{i,v-i}}{a(v)}
  = \frac{2F_{v-i,i}}{a(v)} . 
\end{equation}
Their sum over $i$ is the total number of fragments, which must be 2. Indeed the condition is satisfied:
\begin{equation}
\label{nof-bin}
  \nof 
  = \sum_{i=1}^{v-1} b_{i|v} 
  = \frac{2}{a(v)}\sum_{i=1}^{v-1}F_{v-i,i} = 2 . 
\end{equation}
Equations (\ref{ak-bin}), (\ref{bik-bin}) and (\ref{nof-bin}) demonstrate how to obtain the fragmentation rate $a(v)$ and fragment distribution $b_{i|v}$ from the fragmentation kernel.

%--------------------------------------------------------------------------*
\subsection{Special Case: Random fragmentation}
A cluster with mass $v$ produces $v-1$ ordered pairs of fragments. If we require all pairs produced by mass $v$ to be equally probable, then we must have
\begin{equation}
\label{F_rnd_1}
  F_{v-i,i} = \mathrm{const.} = C_v
\end{equation}
for all $i$. If we further require that pairs be equally probable regardless of the parent mass that produced them, then $C_v$ must be the same for all $v$. Setting $C_v=1$ we obtain the kernel of random binary fragmentation:
\begin{equation}
\label{F_rnd_2}
  F_{v-i,i} = 1 . 
\end{equation}
This kernel ensures randomness on two separate levels: (i) all ordered pairs of fragments produced by the same parent are equally probable, (ii) also equally probable relative to pairs produced by any other particle.  Condition (i) is set by Eq.\ (\ref{F_rnd_1}) and condition (ii) by the additional requirement in (\ref{F_rnd_2}). These conditions are independent: it is possible to construct kernels that satisfy the first condition but not the second: simply set $C_v$ to some function of $v$ other than $1$. 
The fragmentation rate follows from Eq.\ (\ref{ak-bin}),
\begin{equation}
  a(v) = \sum_{i=1}^{v-1}(1) = v-1,
\end{equation}
and the distribution of fragments from Eq.\ (\ref{bik-bin}),
\begin{equation}
  b_{i|v} = 2/v .
\end{equation}
Accordingly, the fragmentation rate is proportional to the number of ordered pairs of fragments from mass $v$ and the distribution of fragments is uniform. 

%--------------------------------------------------------------------------*
\section{The kernel in multinary fragmentation}
The binary case suggests a general interpretation of the kernel that is independent of the number of fragments: the fragmentation kernel is the rate of formation of an ordered sequence of fragments from a given mass. It is a function of the masses of all fragments produced and has the following properties: it is invariant to permutations of the masses of the fragments; it is proportional to the probability to obtain the particular ordered sequence of fragments, and the proportionality constant is the fragmentation rate of the parent mass. 
To express this relationship in concise form we use  the vector notation $\bm{v}=(v_1,v_2\cdots v_\nof$), to indicate the ordered list of fragments from mass $v$, where $v_i$ are $\nof$ non zero integers that satisfy
\begin{equation}
  \sum_i   v_i = \nof,\quad
  \sum_i i v_i = v  . 
\end{equation}
In this notation the aggregation kernel is
\begin{equation}
\label{kernel:def}
  F(\bm{v}) = a(v)\, \mathrm{Prob}(\bm{v}) . 
\end{equation}
We will discuss possible ways to assign probabilities to configurations but for now we assume that these probabilities are given and satisfy two conditions: they are properly normalized over all possible fragment distributions from the same parent particle, and invariant to permutations of the order of fragment masses. The latter ensures that the kernel itself is invariant to permutations in the order of the fragments.  The normalization condition is
\begin{equation}
\label{prob_normalization}
  \sum_{\bm{v}|v}  \mathrm{Prob}(\bm{v}) = 1,
\end{equation}
with the summation taken over all ordered sequences of $\nof$ fragments from parent mass $v$. This domain contains all ordered partitions of  $v$ into $\nof$ part. By straightforward combinatorial enumeration the number of configurations in the domain defined by $v$ and $\nof$ is \cite{Matsoukas:book:2019,Matsoukas:CondMatter:2020} 
\begin{equation}
\label{Omega_rnd_1C}
  \Omega^*_{v;\nof} = \binomial{v-1}{\nof-1}. 
\end{equation}
Summations over fragment configurations $\bm{v}$,  as in Eq.\ (\ref{prob_normalization}), are understood to be over the $\Omega^*_{v;\nof}$ elements of this space. 

We now examine the properties of the kernel defined in Eq.\ (\ref{kernel:def}). 
The fragmentation rate follows by summing Eq.\ (\ref{kernel:def}) over all configurations:
\begin{equation}
\label{K:av}
   a(v)  = \sum_{\bm{v}|v} F(\bm{v}).
\end{equation}
Substituting the result into Eq.\ (\ref{kernel:def}) we obtain the probability of configuration,
\begin{equation}
\label{K:prob}
   \mathrm{Prob}(\bm{v}) 
   = \frac{F(\bm{v})}{\sum_{\bm{v}|v} F(\bm{v})} . 
\end{equation}
To obtain the mean number of fragments we note that configuration $\bm{v}$ contributes $b_{i|v}$ fragments of size $i$ with probability $ \mathrm{Prob}(\bm{v})$. Its mean over the ensemble of all fragment configurations is
\begin{equation}
\label{K:biv}
  \bar b_{i|v} 
  = \sum_{\bm{v}|v} b_{i|v}\,  \mathrm{Prob}(\bm{v})
  = \frac{1}{a(v)}\sum_{\bm{v}|v} b_{i|v}\, F(\bm{v}) .
\end{equation}
It is straightforward to show that the sum of all $\bar b_{i|v}$ is $\nof$, as it should:
\begin{equation}
\label{K:B}
  \sum_i \bar b_{i|v}
  % = \sum_i \sum_{\bm{v}|v}  b_{i|v}\,  \mathrm{Prob}(\bm{v})
  = \sum_{\bm{v}|v} \,  \mathrm{Prob}(\bm{v})  \sum_i b_{i|v}
  =  \nof \sum_{\bm{v}|v} \,  \mathrm{Prob}(\bm{v}) 
  = \nof.
\end{equation}
These results are summarized in Table \ref{tbl_kernel} and show how  all properties of fragmentation are related to the kernel.

%--------------------------------------------------------------------------* 
\begin{table}
\caption{Fragmentation Kernel in Multinary One-Component Fragmentation}  
\label{tbl_kernel}
\begin{equation*}
\renewcommand{\arraystretch}{2}
\begin{array}{ *2{ >{\displaystyle}l} }
\hline
  \mathrm{Definition} & 
  F(\bm{v}) = a(v)\,  \mathrm{Prob}(\bm{v})
  \\ 
  \textrm{Probability of configuration~~~} & 
   \mathrm{Prob}(\bm{v}) 
  = \frac{F(\bm{v})}{\sum_{\bm{v}|v} F(\bm{v})}
  \\ 
  \textrm{Breakage Rate} &
  a(v)  = \sum_{\bm{v}|v} F(\bm{v})
  \\ 
  \textrm{Fragment Distribution} &
   \bar b_{i|v} 
  = \frac{1}{a(v)}\sum_{\bm{v}|v} b_{i|v}\, F(\bm{v})
  \\ 
  \hline
\end{array}   
\end{equation*}
\end{table}
%--------------------------------------------------------------------------* 

%--------------------------------------------------------------------------* 
\subsection{The fragmentation equation}
The population balance equation is constructed by expressing the formation and depletion rates of particle size $v$ in terms of the fragmentation kernel. Using $c(v)$ for the number of particles with mass $v$ we write:
\begin{equation}
\label{pbe1}
  \frac{\partial c(v)}{\partial t}
  =
  -\sum_{\bm{v}|v} F(\bm{v}) c(v) 
  +\sum_{v'>v} \sum_{\bm{v'}|v'} b_{v|v'} F(\bm{v}') c(v') .
\end{equation}
The first term on the right-hand side is the the rate of fragmentation of mass $v$ with the summation over all $\nof$-tuples generated from that mass. The second term is the production rate of size $v$ with the summation in $\bm{v}'|v'$ over all fragment configurations from mass $v'$ and the summation in $v'>v$ over all particles larger than $v$.  
Applying Eq.\ (\ref{K:av}) to the first summation on the right-hand side of Eq.\ (\ref{pbe1}) and Eq.\  (\ref{K:biv}) to the second summation the fragmentation equation becomes
\begin{equation}
\label{pbe2}
  \frac{\partial c(v)}{\partial t}
  =
  -a(v) c(v) 
  +\sum_{v'>v} \bar b_{v|v'} a(v') c(v') .
\end{equation}
This is the more familiar form of the population balance expressed in terms of the breakage rate $a(v)$ and the mean distribution of fragments $\bar b_{i|v}$. 
%--------------------------------------------------------------------------* 
\subsection{Random fragmentation}
\label{sct_rnd_1C}
As in the binary case we define random multinary fragmentation by the condition that all fragment configurations be produced at the same rate regardless of the particle that produces them; thus we require
\begin{equation}
\label{rnd_Kv}
  F(\bm{v}) = 1 , 
\end{equation}
for all $\bm{v}$. The probability of configuration is uniform,
\begin{equation}
  \label{rnd_Probv}
  \mathrm{Prob}(\bm{v}) = 1/\Omega^*_{v;\nof} ,
\end{equation}
where $\Omega^*_{v;\nof}$ is the total number of configurations produced by mass $v$ and is given by Eq.\ (\ref{Omega_rnd_1C}). 
The fragmentation rate follows from Eq.\ (\ref{K:av}):
\begin{equation}
\label{rnd_av}
  a(\bm{v}) = \Omega^*_{v;\nof} .
\end{equation}
The mean distribution  of fragments is given by Eq.\ (\ref{K:biv}) but in this case follows from a special result of the ensemble of equiprobable partitions \cite{Matsoukas:book:2019,Matsoukas:CondMatter:2020} 
\begin{equation}
\label{rnd_b:mean}
  \bar b_{i|v} = \nof \frac{\Omega^*_{v-i;\nof-1}}{\Omega^*_{v;\nof}}
  \equiv \bar b^*_{i|v} . 
\end{equation}
Equations (\ref{rnd_Kv})--(\ref{rnd_b:mean}) summarize the results for random $\nof$-nary fragmentation. 
Below we give the explicit forms for binary and ternary fragmentation kernels; higher order kernels may be obtained similarly. 

%--------------------------------------------------------------------------* 
\paragraph{Example:  Binary Random Fragmentation} With $\nof=2$, $v=i+j$, we obtain
\begin{equation}
  \begin{array}{>{\displaystyle}l}
    F_{i,j} = 1 , \\
    a(v) = \Omega^*_{v;2} = v-1 ,\\ 
    \mathrm{Prob}(i,j) 
      = \frac{1}{\Omega^*_{v;2}} = \frac{1}{v-1} \\ 
    \bar b^*_{i|v} = 2\frac{\Omega^*_{v-i;1}}{\Omega^*_{v;2}} , 
      =  \frac{2}{v-1} . 
  \end{array}
\end{equation}
Thus we recover the results of random binary fragmentation. 

%--------------------------------------------------------------------------* 
\paragraph{Example: Ternary Random Fragmentation}  With $\nof=3$, $v=i+j+k$, we have
\begin{equation}
  \begin{array}{>{\displaystyle}l}
      F_{i,j,k} = 1 \\
      a(v) = \Omega^*_{v;3} = \frac{1}{2} (v-2) (v-1) \\ 
      \mathrm{Prob}(i,j,k) = \frac{1}{\Omega^*_{v;3}} = \frac{2}{(v-2)(v-1)} \\ 
      \bar b_{i|v} = 3 \frac{\Omega^*_{v-i;2}}{\Omega^*_{v;3}}
    = 6 \frac{v-i-2}{(v-2) (v-1)} .
  \end{array}
\end{equation}

Unlike the binary case the distribution of fragments is not uniform because the conditions of constant $v$ and $\nof$ on the distribution of fragments introduce correlations between the size of fragments. For large $\nof$ and $v\gg n$ the distribution in random fragmentation becomes exponential with mean fragment size $v/\nof$ \cite{Matsoukas:Entropy:2022}.

%--------------------------------------------------------------------------* 
\subsection{Partially random fragmentation -- kernel depends only on the total mass}

A first level of deviation from random fragmentation is when the mean distribution of fragments is given by Eq.\ (\ref{rnd_b:mean}) as in random fragmentation, but the breakage rate is arbitrary. A second level of deviation is when both the distribution of fragments and the breakage rate are arbitrary.
The former case, which we call partially random fragmentation, is  the simplest type of deviation from the fully random case. Under partially random fragmentation we have:
\begin{equation}
\label{partially_rnd_1C_F}
\begin{array}{l}
  F_{\bm{v}} = \kappa(v);\quad  
  a(\bm{v})  = \kappa(v) \Omega^*_{v;\nof};\quad
   \mathrm{Prob}(\bm{v}) = 1/\Omega^*_{v;\nof};\quad 
  b_{i|k} = b^*_{i|k} . 
\end{array}
\end{equation}
In partially random fragmentation each cluster produces equiprobable configurations but these are weighed unequally depending on the size of the parent cluster that produces them. Certain cases of partially random fragmentation were solved by \cite{Ziff:JPMG85} for the binary case. 

%--------------------------------------------------------------------------* 
\subsection{Arbitrary distribution of fragments}
In the most general case the probability of configuration depends on all masses in the configuration of fragments. We introduce such dependence by setting the probability of configuration $\bm{b}$ proportional to some functional $W(\bm{b})$ of the distribution $\bm{b} = (b_{1|v},b_{2|v}\cdots)$ of the masses in $\bm{v}$. By making $W$ a functional of the distribution rather than a function of the configuration itself we ensure that all permutations within configuration $\bm{v}$ are equally probable as they all have the same distribution. The probability of configuration is
\begin{equation}
\label{prob:W}
   \mathrm{Prob}(\bm{v}) = \frac{W(\bm{b})}{\Omega_{v;\nof}} ,
\end{equation}
where $\Omega_{v;\nof}$ is the partition function of the ensemble of fragments, defined by the normalization condition
\begin{equation}
  \Omega_{v;\nof} 
    = \sum_{\bm{v}|v} W(\bm{b}) . 
\end{equation}
A special class of functionals is of the form
\begin{equation}
\label{W_1C}
  W(\bm{b}) = w_1^{b_1} w_2^{b_2}\cdots
\end{equation}
where $w_i$ depend on $i$ but are the same for all distributions. The mean fragment distribution in this case is \cite{Matsoukas:book:2019}
\begin{equation}
  \bar b_{i|v}  = \nof\,w_i \frac{\Omega_{v-i;\nof-1}}{\Omega_{v;\nof}} .
\end{equation}
Random fragmentation is recovered by setting $w_i=1$.  It is not always possible to obtain the partition function in closed form in terms of $v$ and $\nof$ for any $w_i$, though scattered results can be found in the literature \cite{Matsoukas:CondMatter:2020}. In principle, once the probability of configuration in Eq.\ (\ref{prob:W}) is specified, the kernel can be constructed according to Eq.\ (\ref{kernel:def}) and all properties of fragmentation follow from the results in Table \ref{tbl_kernel}.

%--------------------------------------------------------------------------* 
\section{The multicomponent case}
%--------------------------------------------------------------------------* 
\subsection{The ensemble of configurations}
We now turn to the problem of bicomponent multinary fragmentation. The extension to any number of components is straightforward and will not be discussed here. We consider a system composed of mixed clusters that contain $v_A$ units of component $A$ and $j$ units of component $B$. The state of this cluster is represented by vector $\bm{v} = (v_A,v_B)$ and its total mass is $v = v_A+v_B$. When such cluster breaks it forms $\nof$ bicomponent fragments $\bm{v}_l = (v_{Ai},v_{Bi})$, $i=1,\cdots \nof$ and the configuration of fragments
is are now represented by two-dimensional vector $\bm{V} = \Big(\bm{v}_1,\bm{v}_2\cdots \bm{v}_\nof\Big)$, a matrix with dimensions $\nof\times 2$. Schematically we write
\begin{equation}
  \bm{v} \stackrel{F(\bm{V})}{\longrightarrow} \bm{V} . 
\end{equation}
The domain consists of all ordered lists of fragments produced from parent particle $\bm{v}=(v_A,v_B)$. We construct it by considering the equivalent combinatorial problem: starting with a set of $v_A$ units of component $A$ and $v_B$ units of component $B$ we  distribute them sequentially into $\nof$ distinguishable fragments (`buckets') so that no bucket is empty. A configuration is defined by the number of units of $A$ and $B$ in each bucket and the order in which each bucket was filled. For example, the following configurations from parent particle $\bm{v} = (2,3)$, a pentamer that contains one A $(=\bullet)$ and three B's $(=\circ)$, are counted as distinct:
\begin{equation}
  \begin{array}{l c c c}
    & \textrm{bucket 1}   && \textrm{bucket 2}
    \\\hline
    \bm{V}_1\hspace{20pt}~&
    \bullet\circ{}      && \bullet\circ{}\circ
    \\
    \bm{V}_2\hspace{20pt}~&
    \bullet{}\circ\circ && \bullet\circ{}  
    \\ 
    \bm{V}_3\hspace{20pt}~&
    \circ{}\bullet{}    && \bullet\circ{}\circ
    \\\hline
  \end{array}
\end{equation}
All three configurations of this example contain two fragments each, a dimer $\bm{u}=(1,1)$ with one A and one B, and a trimer $\bm{u'}=(1,2)$ with one A and two B's. Configurations  $\bm{V}_1$ and $\textbf{V}_2$ contain identical clusters  but in different buckets; $\bm{V}_1$ and $\bm{V}_3$ are distinguished because the order of components in bucket 1 is different.   The presence of two components increases the number of configurations relative to the one-component case by the factor
\begin{equation}
\label{omega_V}
  \omega(\bm{V}) = \prod_{\alpha,\beta}
  \binomial{\alpha+\beta}{\alpha}^{b_{\alpha,\beta}} , 
\end{equation}
where $b_{\alpha,\beta}$ is the number of fragments with $\alpha$ units of $A$ and $\beta$ units of $B$ in configuration $\bm{V}$. By combinatorial calculation the number of configurations in $\nof$-nary bicomponent fragmentation  is \cite{Matsoukas:CondMatter:2020} 
\begin{equation}
\label{omega_rnd_2C}
  \Omega^*_{v_A,v_B;\nof} 
  = \sum_{\bm{V}|\bm{v}} (1)
  = \frac{(v_A+v_B)!}{v_A! v_B!}\,\Omega^*_{v_A+v_B;\nof} ,    
\end{equation}
where $\Omega^*_{v_A+v_B;\nof}$ is the number of fragments in one component fragmentation from mass $v_A+v_B$, given in Eq.\ (\ref{Omega_rnd_1C}). The binomial term on the right-hand side expresses the effect of composition, which is to increase the number of fragment configurations by the number of permutations in the order of the components within the fragments of configuration.

%--------------------------------------------------------------------------* 
\subsection{The multicomponent kernel}
By analogy to the one-component case the multicomponent fragmentation kernel $F(\bm{V})$ is the rate of formation of configuration $\bm{V}$ from parent cluster $\bm{v}$:
\begin{equation}
\label{multi_kernel_def}
    F(\bm{V}) = a(\bm{v})  \mathrm{Prob}(\bm{V}) .  
\end{equation}
The probability of configuration, $ \mathrm{Prob}(\bm{V})$, is obtained from the kernel as
\begin{equation}
   \mathrm{Prob}(\bm{V})
  =
  \frac{F(\bm{V})}{\sum_{\bm{V}|\bm{v}} F(\bm{V})}, 
\end{equation}
with the summation over all configurations from parent cluster $\bm{v}$. The fragmentation rate of cluster is
\begin{equation}
  a(\bm{v}) = \sum_{\bm{V}|\bm{v}} F(\bm{V}) .
\end{equation}
Finally, the mean number $\bar b_{\bm{u}|\bm{v}}$ of fragments $\bm{u} = (u_A,u_B)$ from parent cluster $\bm{v} = (v_A,v_B)$ is
\begin{equation}
  \bar b_{\bm{u}|\bm{v}}
  =
  \sum_{\bm{V}|\bm{v}} b_{\bm{u}|\bm{v}}  \mathrm{Prob}(\bm{V})
  =
  \frac{1}{a(\bm{v})}
  \sum_{\bm{V}|\bm{v}} b_{\bm{u}|\bm{v}} F(\bm{\bm{V}}) . 
\end{equation}
The results are summarized in Table \ref{tbl_kernel_multi} and are direct extensions of the corresponding results for the one-component case in Table \ref{tbl_kernel}. 

%--------------------------------------------------------------------------* 
\subsection{The multicomponent/multinary fragmentation equation}
The fragmentation equation is constructed similarly to the one-component case. Denoting the concentration of fragments $\bm{v}$ by $c(\bm{v})$, we write
\begin{equation}
  \frac{\partial c(\bm{v})}{dt}
  =
  -\sum_{\bm{V}|\bm{v}} F(\bm{V}) c(\bm{v})
  +\sum_{\bm{v'}>\bm{v}}
   \sum_{\bm{V'}|\bm{v'}} 
   b_{\bm{v}|\bm{v'}} F(\bm{V'}) c(\bm{v'}) . 
\end{equation}
The first summation on the right-hand side is over all configurations produced by cluster mass $\bm{v}$ and gives the rate of depletion of mass $\bm{v}$. The double summation is the generation of mass $\bm{v}$ via fragmentation of all larger sizes $\bm{v'}>\bm{v}$: the inner summation goes over all configurations produced by mass $\bm{v'}$ and the outer summation over all masses $\bm{v'}>\bm{v}$. 
Using the results in Table \ref{tbl_kernel_multi} this can be written as
\begin{equation}
  \frac{\partial c(\bm{v})}{dt}
  =
  -a(\bm{v}) c(\bm{v})
  + \sum_{\bm{v'}>\bm{v}}
    b_{\bm{v}|\bm{v'}} a(\bm{v'}) c(\bm{v'}) .
\end{equation}
In this form the fragmentation equation is expressed in terms of the breakage rate and the fragment distribution.

%--------------------------------------------------------------------------* 
\begin{table}
\caption{Fragmentation Kernel in Multinary Multicomponent Fragmentation}  
\label{tbl_kernel_multi}
\begin{equation*}
\renewcommand{\arraystretch}{2}
\begin{array}{ *2{ >{\displaystyle}l} }
\hline
  \textrm{Definition} & 
  F(\bm{V}) = a(\bm{v})\,  \mathrm{Prob}(\bm{V})
  \\
  \textrm{Probability of configuration~~~} & 
   \mathrm{Prob}(\bm{V}) 
  = \frac{F(\bm{V})}{\sum_{\bm{V}|\bm{v}} F(\bm{V})}
  \\ 
  \textrm{Breakage Rate} &
  a(\bm{v})  = \sum_{\bm{V}|\bm{v}} F(\bm{V})
  \\ 
  \textrm{Fragment Distribution} &
   \bar b_{\bm{u}|\bm{v}} 
  = \frac{1}{a(\bm{v})}
    \sum_{\bm{V}|\bm{v}} b_{\bm{u}|\bm{v}}\, F(\bm{V})
  \\ 
  \hline
\end{array}   
\end{equation*}
\end{table}
%--------------------------------------------------------------------------* 

%--------------------------------------------------------------------------* 
\subsection{Random multicomponent fragmentation}

As in one component fragmentation, the easiest case to treat is random fragmentation, which we define by the condition that all configurations are equally probable regardless of the size or composition of the parent cluster. Accordingly the kernel has the same value for all configurations regardless of the mass they contain:
\begin{equation}
  F(\bm{V}) =  1. 
\end{equation}
The fragmentation rate of cluster $\bm{v} = (v_A,v_B)$ is equal to the number of configurations produced by that mass, $\Omega_{v_A,v_B;\nof}$ in Eq.\ \ref{omega_rnd_2C}; accordingly, the fragmentation rate is 
\begin{equation}
  a(\bm{v}) 
  = \Omega^*_{v_A,v_B;\nof} 
\end{equation}
and the probability of configuration is 
\begin{equation}
   \mathrm{Prob}(\bm{V}) = 1/\Omega_{v_A,v_B;\nof} . 
\end{equation}
The mean number of fragments of size $\bm{u}=(u_A,u_B)$ produced from parent cluster $\bm{v}=(v_A,v_B)$ when all configurations are equally probable was derived in \cite{Matsoukas:CondMatter:2020}:
%
% Eq 45 in Condensed Matter
%
\begin{equation}
\label{multi_rnd_b:mean}
  \bar b_{\bm{u}|\bm{v}}  
  =
  \bar b^*_{u|v} 
  \left.
  {
   \binomial{v_A}{u_A} \binomial{v_B}{u_B}
  }
  \right/
  {
  \binomial{v_A+v_B}{u_A+u_B}
  } .
\end{equation}
Here $u=u_A+u_B$ is the mass of fragment $\bm{u}$, $v=v_A+v_B$ is the mass of parent $\bm{v}$ and $\bar b^*_{u|v}$ is the mean number of fragments of size $u$ from parent $v$ in random fragmentation for one component from Eq.\ (\ref{rnd_b:mean}). Equation  (\ref{multi_rnd_b:mean}) expresses the mean distribution as a product of two distributions: the fragment distribution in one component random fragmentation ($\bar b^*_{u|v}$) and a factor of binomials that represent the compositional distribution of the fragments. 

%--------------------------------------------------------------------------*
\subsection{Partially random fragmentation}
This case follows by analogy of the one-component case in Eqs.\ (\ref{partially_rnd_1C_F}): 
\begin{equation}
  \begin{array}{>{\displaystyle}l}
    F(\bm{V}) = \kappa(v),\quad   
    a(\bm{V}) = \kappa(v) \Omega^*_{v_A,v_B;\nof},\quad  
    \mathrm{Prob}(\bm{V}) = 1/ \Omega^*_{v_A,v_B;\nof}, \\ 
    \bar b_{\bm{u}|\bm{v}} =
      \bar b^*_{u|v} 
      \left.
      {\binomial{v_A}{u_A} \binomial{v_B}{u_B}}\right/
      {\binomial{v_A+v_B}{u_A+u_B}} .
  \end{array}
\end{equation}
In partially random fragmentation the kernel is a function of the total mass in the configuration of fragments but not of the masses of the fragments or their composition; the probability of configuration and the mean number of fragments is the same as in random fragmentation but the breakage rate varies from that in random fragmentation by a factor $F(v)$. 

%--------------------------------------------------------------------------*
\subsection{A case of non random fragmentation}\label{sct_non-rnd_2C}
We consider now a case of non-random fragmentation where individual components break up independently of each other. The process works as follows: particle $\bm{v} = (v_A,v_B)$ produces $\nof$ fragments of component $A$ and $nof$ fragments of component $B$ which then combine to form $\nof$ pairs of $A$-$B$ fragments. We will construct the kernel under the condition that fragmentation of each component and the recombination of fragment pairs are both random, i.e., they produce every possible outcome with equal probability. Cluster mass $v$ breaks with rate $a(v)$ for one-component random fragmentation,  Eq.\ (\ref{rnd_av}). Independence implies that the fragmentation rate is the product of the rate of each component:
\begin{equation}
  a(\bm{v}) = a(v_A) a(v_B) = \Omega^*_{v_A;\nof} \Omega^*_{v_B;\nof} . 
\end{equation}
The breakup of component mass $v_A$ produces $\Omega^*_{v_A,\nof}$ configurations of $\nof$ fragments, and similarly for component $B$. Configurations of the two components combine in all possible ways to produce a total of $\Omega^*_{v_A;\nof} \Omega^*_{v_B;\nof}$ bicomponent configurations all of which appear with the same probability. The pairing of $\nof$ fragments of $A$ with $\nof$ fragments of $B$ produces a configuration $\bm{V}$, which may be manifested in $\omega(\bm{V})$ ways in Eq.\  (\ref{omega_V}), corresponding to the order in which the two components appear in each fragment. These compositional permutations are all equally probable. Therefore, the probability of configuration in this model is
\begin{equation}
   \mathrm{Prob}(\bm{V}) 
  = 
  \frac{1}{\omega(\bm{V}) \Omega^*_{v_A;\nof} \, \Omega^*_{v_B;\nof}} . 
\end{equation}
The kernel is obtained by combining the fragmentation rate and the probability of configuration according to Eq.\ (\ref{multi_kernel_def}):
\begin{equation}
  F(\bm{V}) = \frac{1}{\omega(\bm{V}) }
  = 
  1\Big/ \prod_{\alpha,\beta} \binomial{\alpha+\beta}{\alpha}^{b_{\alpha,\beta}} . 
\end{equation}
Here $b_{\alpha,\beta}$ is the number of fragments in configuration $\bm{V}$ that contain $a$ units of $A$ and $\nof$ units of $B$  and $\omega(\bm{V})$ is the number of permutations in the order in which components are assigned to each fragment (Eq.\ \ref{omega_V}). This of the form
\begin{equation}
  F(\bm{V}) = \frac{W(\bm{b}|\bm{V})}{\Omega_{v_A,v_B;\nof}} , 
\end{equation}
where $\Omega_{v_A,v_B;\nof}$ is the partition function and $W(\bm{b}|\bm{V})$ is a functional of the distribution of fragments,
\begin{equation}
  W(\bm{b}|\bm{V}) 
  = \prod_{\alpha,\beta} w_{a,b}^{b_{\alpha,\beta}}
  = \prod_{\alpha,\beta} \binomial{\alpha+\beta}{\alpha}^{-b_{\alpha,\beta}} . 
\end{equation}
This is an extension  of Eq.\ (\ref{W_1C}) to non random fragmentation of multicomponent clusters. In this model configurations are undersampled relative to the random case by a factor equal to the number of permutations in the order of the components in the distribution of fragments. 

%--------------------------------------------------------------------------*
\section{Concluding Remarks}
The fragmentation kernel arises from the probabilistic nature of fragmentation. The fundamental event is the formation of a discrete set of fragments;  the rate of its formation is the fragmentation kernel. The central relationship is  Eq.\ (\ref{multi_kernel_def}) that links the kernel to the rate of fragmentation as a function of mass and the probability distribution of the fragments. An element of the formulation is the order of fragments and the order of components within fragments. Consideration of the order arises from the convention that specifies the domain of fragment configurations through the requirement that every permutation be visited exactly once. 
 
The two key properties that define the kernel are the fragmentation rate as a function of size and the probability of fragment configurations. The kernel is constructed from these via  Eq.\ (\ref{multi_kernel_def}). The mean number of fragments, which appears in the fragmentation equation, is a reduced order property, a moment of the probability distribution that conveys less information than the kernel itself. This is in turn implies that the kernel contains more information than the fragmentation equation requires. What is then the value of the kernel? In answering this question we must recognize that the fragmentation equation itself is a reduced-order description of fragmentation that gives the mean distribution in the ensemble of all feasible distributions. The formal probabilistic treatment of the fragmentation ensemble assigns a probability to every feasible outcome and obtains the mean distribution by proper averaging over this ensemble \cite{Matsoukas:Entropy:2022}. The failure of the fragmentation equation in shattering \cite{McGrady:PRL87,Ziff:JPMG92,Matsoukas:Entropy:2022} is one indication that this equation conveys incomplete information about the process. 
From a more practical perspective, the path to the construction of a fragmentation model goes through the kernel: if we hypothesize the probability of any fragment configuration and the rate of breakage, the process is fully specified. This is precisely how we obtained the kernels for random fragmentation and for independent fragmentation of components. It is possible that one could construct reduced-order models for the mean number of fragments but the process would remain incompletely specified unless the fragmentation kernel is known.

%--------------------------------------------------------------------------*
\section*{References}
\bibliographystyle{abbrvnat}
\bibliography{StatMech,tm}

\end{document}